\documentclass[submission,
copyright,creativecommons]{eptcs}
\providecommand{\event}{GandALF 2022}
\usepackage{underscore}           

\usepackage[ruled]{algorithm2e}
\newtheorem{definition}{Definition}
\newtheorem{lemma}{Lemma}

\usepackage{amsmath}
\usepackage{amssymb}
\usepackage{amsfonts}
\usepackage{tikz}   
\tikzstyle{tplace}=[circle,draw,inner sep=1.5mm]
\tikzstyle{transition} = [rectangle, draw, inner sep=0,text width=0.4cm, fill=gray!40, minimum height=0.4cm, text centered]
\tikzstyle{state} = [circle, draw, inner sep=0,text width=0.4cm, minimum height=0.4cm, text centered]
\tikzstyle{event} = [rectangle, draw=green!50!black!75,fill=green!50!black!20, inner sep=0,text width=0.7cm,
minimum height=0.5cm, text centered,outer sep=0]
\tikzstyle{cond} = [circle, draw=blue!75,fill=blue!10, inner sep=0,text width=0.4cm,
minimum height=0.4cm, text centered]
\tikzstyle{none} = [draw=none, fill=none, inner sep=0,minimum height=0.2cm]
 \tikzstyle{leer} = [rectangle, draw=none, inner sep=0, text  centered]
\tikzstyle{every label}=[red]
\tikzstyle{edge} = [thick]
\newcommand{\define}{\stackrel{\triangle}{=}}

\newtheorem{theorem}{Theorem}
\newcommand{\unfolding}{\mathcal{U}}

\newcommand{\markis}{\mathcal{M}}
\newcommand{\markn}{\mathit{M}}
\newcommand\marking[1]{\mathit{Mark}(#1)}

\newcommand\badstates{\mathcal{Z}}
\newcommand\clbadstates{\overline{\badstates}}
\newcommand\badconfs{\mathcal{B}}

\newcommand\attn{\mathbf{A}}
\newcommand{\preset}[1]{{^{\bullet}}{#1}}
\newcommand{\postset}[1]{{{#1}^{\bullet}}}
\newcommand{\adequate}{\prec}
\newcommand\spoil[1]{\mathbf{spoil(#1)}}
     
\newcommand{\concurrent}{~\mathbf{co}~}
\newcommand\concat{\oplus}
\newcommand\konkat{\bigoplus}
\newcommand\leftchop{\ominus}
\newcommand\rightchop{\oslash}
\newcommand{\conflict}{\mathrel{\#}}        
  
\newcommand{\dircf}{\conflict_\delta}  
\newcommand{\strcf}{\conflict_\sigma}  
\newcommand\cone[1]{[#1]}

\newcommand{\tup}[1]{\langle {#1} \rangle}
\newcommand\onlynetn{{\mathit{N}}}
\newcommand{\net}{\onlynetn}

\newcommand\netn{\mathcal{N}}
\newcommand\petrinet{\netn}
\newcommand{\places}{\mathit{P}}
\newcommand{\place}{\places}
\newcommand\transn{\mathit{t}}
\newcommand\onet{\mathcal{O}}
\newcommand\trans{{\mathit{T}}}
\newcommand\placen{\mathit{p}}
\newcommand\flow{\mathit{F}}
\newcommand\oflow{\mathit{G}}

\newcommand{\edges}{\mathcal{E}}

\newcommand\prefix{\preceq}
\newcommand\prefixn{\Pi}

\newcommand{\prefn}{\Pi}
\newcommand\by{\begin{eqnarray}}
\newcommand\ey{\end{eqnarray}}
\newcommand\bys{\begin{eqnarray*}}
\newcommand\eys{\end{eqnarray*}}
\newcommand\bdf{\begin{Definition}}
\newcommand\edf{\end{Definition}}
\newcommand\bet{\begin{theorem}}
\newcommand\ent{\end{theorem}}
\newcommand\bel{\begin{lemma}}
\newcommand\bec{\begin{corollary}}
\newcommand\enc{\end{corollary}}
\newcommand\bum{\begin{enumerate}}
\newcommand\eum{\end{enumerate}}
\newcommand\bit{\begin{itemize}}
\newcommand\eit{\end{itemize}}
\newcommand\bepr{{\bf Proof:} \ \nopagebreak}
\newcommand\eepr{\hspace{\fill} \nolinebreak $\Box$}

\newcommand\NN{\mathbb{N}}

\newcommand\oplace{\mathit{B}}
\newcommand\otrans{\mathit{E}}
\newcommand\oplacen{\mathit{b}}
\newcommand\otransn{\mathit{e}}

\newcommand\runn{\omega}
\newcommand\runs{\Omega}
\newcommand{\states}{\mathcal{M}}
\newcommand\move[1]{\stackrel{#1}{\rightarrow}}
\newcommand\setgdw{\stackrel{\triangle}{\Longleftrightarrow}}
\newcommand\omove[1]{\stackrel{#1}{\leadsto }}

\newcommand\reach{\mathbf{R}}

\newcommand\confign{\mathit{C}}

\newcommand\configs{\mathcal{C}}
\newcommand\finfigs{\mathcal\configs^{\mathbf{f}}}
 
 \newcommand\doofigs{\mathcal{D}}
 \newcommand\viables{\mathcal{F}}
\newcommand\mindofigs{\check{\doofigs}}
 \newcommand\mindooout{\mathfrak{D}}

\newcommand\cutn{\mathbf{c}}
\newcommand\cut{\mathbf{cut}}

\newcommand\trunk[1]{\langle#1\rangle}
\newcommand\crest[1]{\mathbf{crest}(#1)}
\newcommand\shaved[1]{\mathbf{shave}(#1)}

\newcommand{\dheight}[1]{\mathbf{dech}(#1)}
\newcommand{\schutz}[1]{\mathbf{prot}(#1)}

\newcommand\watershed{\gamma}
\newcommand\tips{\Gamma}
\newcommand\ridge{\chi}

\newcommand{\worklist}{\mathsf{wl}}
\newcommand\addit{\mathsf{add}}
\newcommand\true{\mathsf{true}}
\newcommand\false{\mathsf{false}}
\newcommand\myleq{\leqslant}

\newcommand\fold{\pi}

\usepackage{mathtools}
\usepackage[draft]{fixme}
\usepackage{epsfig}
\usetikzlibrary{arrows,shapes,decorations,automata,backgrounds,petri,positioning}
\usepackage{amsfonts}
\usepackage{amsmath}
\usepackage{graphicx}      
\usepackage{caption}
\usepackage{subcaption}

\SetKwComment{Comment}{/* }{ */}
\def\titlerunning{Finding Minimal Doomed Configurations}
\def\authorrunning{Aguirre-Sambon\'i et al.}

\title{Avoid One's Doom: Finding Cliff-Edge Configurations  in Petri Nets
\author{Giann Karlo Aguirre-Sambon\'i\thanks{
  We gratefully acknowledge the fruitful exchanges with C\'edric Gaucherel and Franck Pommereau. This work was supported by the \emph{DIGICOSME} grant \textsc{Escape}, \emph{DIGICOSME RD 242-ESCAPE-15203}, and by the French Agence Nationale pour la Recherche (ANR) in the scope of the
project ``BNeDiction'' (grant number ANR-20-CE45-0001).
}$\;\;^1$\qquad
Stefan Haar\footnotemark[1]$\;\;^1$ \qquad
Lo\"ic Paulev\'e\footnotemark[1]$\;\;^2$ \\
Stefan Schwoon\footnotemark[1]$\;\;^1$ \qquad
Nick W\"urdemann\footnotemark[1]$\;\;^3$
\institute{$^1$INRIA and LMF, CNRS and ENS Paris-Saclay, Universit\'e Paris-Saclay, Gif-sur-Yvette, France \email{$\{$giann-karlo.aguirre-samboni,stefan.haar,stefan.schwoon$\}$@inria.fr}
}
\institute{$^2$Univ. Bordeaux, Bordeaux INP, CNRS, LaBRI, UMR5800, Talence, France
\email{loic.pauleve@labri.fr}}
\institute{$^3$Department of Computing Science, University of Oldenburg, Oldenburg, Germany 
\email{wuerdemann@informatik.uni-oldenburg.de}}}}
\begin{document}
\maketitle
\begin{abstract}
A crucial question in analyzing a concurrent system is to determine its long-run behaviour, and in particular, whether there are irreversible choices in its evolution, leading into parts of the reachability space from which there is no return to other parts. Casting this problem in the unifying framework of safe Petri nets, our previous work~\cite{CHJPS-cmsb14} has provided techniques for identifying \emph{attractors}, i.e.\ terminal strongly connected components of the reachability space, whose attraction basins we wish to determine. Here, we provide a solution for the case of safe Petri nets.
Our algorithm uses net unfoldings and provides a map of all of the system's configurations  (concurrent executions)  that act as \emph{cliff-edges}, i.e.\ any maximal extension for those configurations lies in some basin that is considered fatal. The computation turns out to require only a relatively small prefix of the unfolding, just twice the depth of Esparza's complete prefix.
\end{abstract}

\section{Introduction}

\emph{Unfoldings} of Petri nets \cite{Esparza08}, which are essentially event structures in the sense of Winskel et al.~\cite{NPW80} with additional information about \emph{states}, %
are an acyclic representation of the possible sequences of transitions, akin to Mazurkiewicz traces 
but enriched with branching information.

Many reachability-related verification problems for concurrent systems have been successfully addressed by Petri-net unfolding methods over the past decades, see~\cite{McM92,ERV02,Esparza08}.
However, questions of long-term behaviour and stabilization have received relatively little attention.  With the growing interest in formal methods for biology, the key feature of  \emph{multistability} of systems~\cite{ta90,Plahte1995,Ozbudak2004,Pisarchik2014} comes into focus. It has been studied in other
qualitative models such as Boolean and multivalued networks~\cite{t80,tt95,Richard2019}.
Multistability characterizes many fundamental biological processes, such as cellular differentiation, cellular reprogramming, and cell-fate decision; in fact, stabilization of a cell regulatory network corresponds to reaching one of the - possibly many - phenotypes of the cell, thus explaining the important role of multistability in cell biology. However, multistability emerges also in  many other branches of the life sciences; our own motivation is  the qualitative analysis of the fate of \emph{ecosystems}, see~\cite{pommereau2022}.

Multistability can be succinctly described as the presence of several \emph{attractors} in the system under study.
Attractors characterize the stable behaviours, given as the smallest subsets of states from which the system cannot escape;  in other words, they are {terminal strongly connected components} of the associated transition system. In the long run, the system will enter one of its attractors and remain inside; multi-stability arises when there is more than one such attractor.
The \emph{basin} of attractor $\attn$ consists of  the states from  which the system inevitably reaches $\attn$. 

The  basin includes the attractor itself, and possibly one or several transient states~\cite{Klarner18}.

We aim at finding the tipping points in which the system switches from an undetermined or \emph{free} state into some basin; while interesting beyond that domain, this is a recurrent question in the analysis of 
 signalling and gene regulatory networks~\cite{Cohen2015,Mendes2018}.
In \cite{bifurcations-BMC}, the authors provide a method for identifying, in a boolean network model, the states in which one transition
leads to losing the reachability of a given attractor (called \emph{bifurcation transitions} there; we prefer to speak of \emph{tipping points} instead).
However, enumerating the states in which the identified transitions make the system branch away from the attractor can be highly combinatorial and hinders a fine understanding of the branching.
Thus, the challenge resides in identifying the specific contexts and sequences of transitions leading to a strong basin.

 Using a bounded unfolding prefix, all reachable  attractors~\cite{CHJPS-cmsb14} can be extracted. Also, we have exhibited (\cite{HPS-cmsb20}) the particular shape of basins that are visible in a concurrent model.

In the present paper, we build on this previous analyses; 
the point of view taken here is that all attractors correspond to the \emph{end} of the system's free behaviour, in other words to its \emph{doom}. We will give characterizations of  basin boundaries (called \emph{cliff-edges} below), and  of those behaviours that remain \emph{free}, in terms of properties of the unfolding, reporting also on practical experiments with an implementation of the algorithms derived.  We finally introduce a novel type of quantitative measure, called \emph{protectedness}, 
to indicate how far away (or close) a system is from doom, in a state that is still free per se. General discussions and outlook will conclude this paper.

\section{Petri Nets and Unfoldings}
\label{sec:unf}
We begin now by recalling the basic definitions needed below.
A \textbf{Petri net} is a bipartite directed graph whose nodes are either \emph{places} or \emph{transitions}, and places may carry \emph{tokens}.
In this paper, we consider only \emph{safe} Petri nets where  a place carries either one or no token in any reachable marking. 
The set of currently active places form the state, or \emph{marking}, of the net.

\textbf{Note.} Some remarks are in order concerning our use of Petri nets versus that of  \emph{boolean networks}, which are more widely used in systems biology. 
Safe (or 1-bounded) Petri nets~\cite{Mur89} are close to Boolean and multivalued networks~\cite{CHKPT-nc19}, yet  enable a more fine-grained specification of the conditions for triggering value changes. Focussing on safe PNs entails no limitation of generality of the model, as two-way behaviour-preserving translations between Boolean and multivalued models exist (see~\cite{CHKPT-nc19} and the appendix of~\cite{CHJPS-cmsb14} for discussion). We are thus entitled to move between these models without loss of expressiveness; however, Petri nets provide more convenient ways to develop and present the theory and the algorithms here.

Formally, a \emph{net} is a tuple $\net=\tup{\place,\trans,\flow}$,  where  $\trans$ is a finite set of  \emph{transitions}, $\place$ a finite set of \emph{places}, 
  and  $\flow\subseteq(\place\times \trans)\cup(\trans\times \place)$ is a \emph{flow relation} whose elements are called \emph{arcs}. 
  In figures,  places are represented by circles and the transitions by boxes (each one with a label identifying it).

For any node $x\in \place\cup \trans$, we call \emph{pre-set} of $x$ the set $\preset{x}=\{y\in \place\cup \trans\mid \tup{y,x}\in \flow\}$ and \emph{post-set} of $x$ the set $\postset{x}=\{y\in \place\cup \trans\mid \tup{x,y}\in \flow\}$.
A subset $\markn\subseteq \place$ of the places is called a \emph{marking}. A \emph{Petri net} is a tuple $\petrinet=\tup{\place,\trans,\flow,\markn_0}$, with $\markn_0\subseteq\place$ an \emph{initial marking}.
Markings are represented by dots (or tokens) in the marked places.
A transition $\transn\in \trans$ is \emph{enabled} at a marking $\markn$, denoted $\markn\move{\transn}$, if and only if $\preset{\transn}\subseteq \markn$. An enabled transition $\transn$ can \emph{fire}, leading to the new marking $\markn'=(\markn\setminus\preset{\transn})\cup\postset{\transn}$;\footnote{This definition does not correspond to the standard semantics of Petri nets, but is equivalent for safe Petri nets, and we prefer it for the sake of simplicity.} in that case we write $\markn\move{\transn}\markn'$.
A \emph{firing sequence} from a marking $M'_0$ is a (finite or infinite) sequence $w=t_1t_2t_3\dots$ over $\trans$ such that there exist markings $\markn'_1,\markn'_2,\dots$ with $\markn'_0\move{t_1}\markn'_1\move{t_2}\markn'_2\move{t_3}\dots$. If $w$ is finite and of length $n$, we write $\markn'_0\move{w}\markn'_n$, and we say that $\markn'_n$ is \emph{reachable} from $\markn'_0$, also simply written $\markn'_0\rightarrow\markn'_n$. We denote the set of markings reachable from some marking $\markn$ in a net $\net$ by $\reach_\net(\markn)$.
A Petri net $\tup{\net,\markn_0}$ is considered \emph{safe} if every marking in $\markn\in\reach_\net(\markn_0)$ and every transition $t$ enabled in $M$ satisfy $(\markn\cap\postset{\transn})\subseteq\preset{\transn}$. In this paper, we assume that all our Petri nets are safe.

From an initial marking of the net, one can recursively derive all possible
transitions and reachable markings, resulting in a \emph{marking graph}
(Def.~\ref{def:marking-graph}).

\begin{definition}
\label{def:marking-graph}
Let $\net=\tup{\place,\trans,\flow}$ be a net and $\states$ a set of markings. The \emph{marking graph} induced by $\states$ is
a directed graph $\tup{\states,\edges}$ such that $\edges\subseteq \states\times \states$ contains $\tup{\markn,\markn'}$ iff $\markn\move{t}\markn'$ for some $\transn\in \trans$; 
the arc $\tup{\markn,\markn'}$ is then \emph{labeled} by $\transn$.
The \emph{reachability graph} of a Petri net $\tup{\net,M_0}$ is the graph induced by $\reach_\net(\markn_0)$.
\end{definition}
The reachability graph is always finite for safe Petri nets.

\begin{figure}[ht]
    \centering
    \begin{subfigure}[t]{0.48\textwidth}
    \def\a{1}
\def\b{0.45}
\def\c{0.7}
\begin{tikzpicture}[>=stealth,shorten >=1pt,node distance=\a cm,auto]
  \node[cond] (p1) at (0, 0)  [label=left:\text{$p_1$}] {$\bullet$};
  \node[cond] (p2) at (6*\b, 0)  [label=left:\text{$p_2$}] {$\bullet$};
  \node[cond] (p3) at (-2*\b, -2*\c)  [label=left:\text{$p_3$}] {};
  \node[cond] (p4) at (2*\b, -2*\c)  [label=left:\text{$p_4$}] {};
  \node[cond] (p5) at (4*\b, -2*\c)  [label=right:\text{$p_5$}] {};
  \node[cond] (p6) at (8*\b, -2*\c)  [label=left:\text{$p_6$}] {};
  \node[cond] (p7) at (6*\b, -5*\c)  [label=below:\text{$p_7$}] {};
  \node[cond] (p8) at (0*\b, -5*\c)  [label=below:\text{$p_8$}] {};
  
  \node[transition] (a) at (-1*\b, -\c) {$\alpha$};
  \node[transition] (b) at (1*\b, -\c) {$\beta$};

  \node[transition] (x) at (-1*\b, -4*\c) {$\xi$};
\node[transition] (c) at (5*\b, -\c) {$\gamma$};
\node[transition] (d) at (7*\b, -\c) {$\delta$};
\node[transition] (e) at (7*\b, -4*\c) {$\theta$};
\node[transition] (r) at (10.5*\b, -1*\c) {$\kappa$};
\node[transition] (f) at (4.5*\b, -4*\c) {$\eta$};
\node[transition] (g) at (1.5*\b, -4*\c) {$\zeta$};
 \path[->] (p1)  edge (a)
        (p3)  edge (x)
        (p3)  edge [bend left=10]    (e)
         (p1)  edge (b)
         (p2) edge (c)
         (p2) edge (d)
         (c) edge (p5)
        (p5) edge (g)
        (p5) edge [bend right=10] (x)
         (p4) edge (g)
         (p4) edge (f)
        (p6) edge (f)
          ; 
 \path[->] (p7) edge [bend right=40] (r)
           (r) edge [bend right=20] (p2) 
           (r) edge [bend right=40] (p1)
 ;
  \path[->]  (a) edge (p3)
  (x) edge (p8)
   (b) edge (p4)
    (d) edge (p6)
    (p2) edge (c)
(p6) edge  (e)
 (e) edge (p7)
 (f) edge (p8)
  (g) edge (p7)  
 ;
 \node at (-1.5*\a,0.45*\a) {(a)\phantomsubcaption\label{fig:basinexample}};
\end{tikzpicture}
    \end{subfigure}
    \begin{subfigure}[t]{0.48\textwidth}
    \def\xdis{1.6cm}
\def\ydis{1.45cm}
\begin{tikzpicture}[>=stealth,shorten >=1pt,node distance=\ydis and \xdis,on grid]
	\node (12) {$ \{p_1, p_2 \} $};
	\node[below =of 12,xshift=-0.5*\xdis] (42) {$ \{p_4, p_2 \} $};
	\node[left=of 42] (32) {$ \{p_3, p_2 \} $};
	\node[below=of 12,xshift=0.5*\xdis] (15) {$ \{p_1, p_5 \} $};
	\node[right=of 15] (16) {$ \{p_1, p_6 \} $};
	\node[below=of 32] (35) {$ \{p_3, p_5 \} $};
	\node[below=of 42] (46) {$ \{p_4, p_6 \} $};
	\node[below=of 15] (36) {$ \{p_3, p_6 \} $};
	\node[below=of 16] (45) {$ \{p_4, p_5 \} $};
	\node[below=of 46,xshift=-0.5*\xdis] (8) {$ \{p_8 \} $};
	\node[below=of 45,xshift=-0.5*\xdis] (7) {$ \{p_7 \} $};
	
	\path[->]
	(12) 	edge node[above left, inner sep= 0.5mm] {\footnotesize$\alpha$} (32)
			edge node[left, inner sep= 1mm] {\footnotesize$\beta$} (42)
			edge node[right, inner sep= 1mm] {\footnotesize$\gamma$} (15)
			edge node[above right, inner sep= 0.5mm] {\footnotesize$\delta$} (16)
	(32)	edge (35)
			edge (36)
	(42)	edge (45)
			edge (46)
	(15)	edge[out=-145,in=50] (35)
			edge (45)
	(16)	edge (36)
			edge[out=-145,in=50] (46)
	(35)	edge node[left, inner sep= 1mm] {\footnotesize$\xi$} (8)
	(46)	edge node[right, inner sep= 1mm] {\footnotesize$\eta$} (8)
	(36)	edge node[left, inner sep= 1mm] {\footnotesize$\theta$}  (7)
	(45)	edge node[right, inner sep= 1mm] {\footnotesize$\zeta$} (7)
	(7)		edge[out=20, in=0, looseness=1.7] node[right, inner sep= 1mm] {\footnotesize$\kappa$} (12)
	;
 \node at (-3,0) {(b)\phantomsubcaption\label{fig:basinstategr}};
\end{tikzpicture}
    \end{subfigure}
    \captionsetup{subrefformat=parens}
    \caption{Petri net example from~\cite{HPS-cmsb20} in \subref{fig:basinexample}, and its reachability graph in \subref{fig:basinstategr}.
    \label{fig:teamplay}
    }
\end{figure}
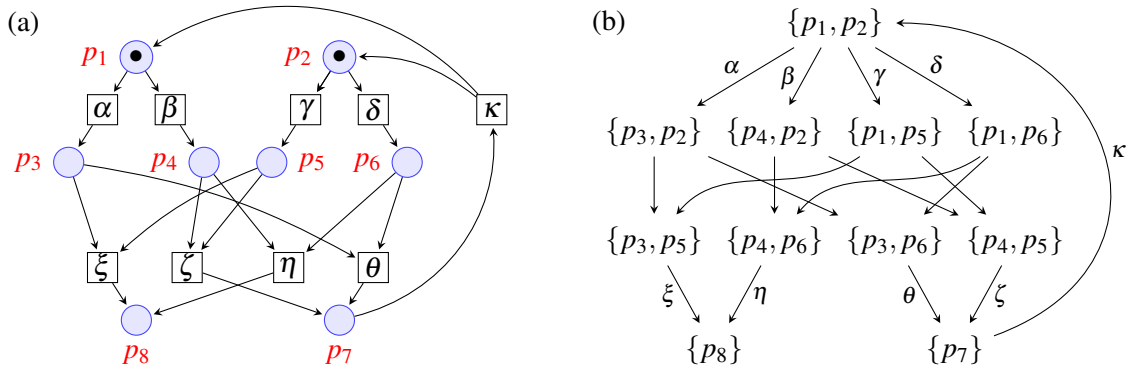
Figure~\ref{fig:basinstategr} shows the reachability graph for our running example \ref{fig:basinexample}.
\\~\\
\textbf{Unfoldings.}
Roughly speaking, the unfolding of
a Petri net $\petrinet$ is an acyclic Petri net (with particular structural properties), $\unfolding$, that reproduces exactly the
same behaviours as $\petrinet$. 

%

We now give some technical definitions to introduce unfoldings formally. A  more extensive treatment  can be found, e.g., in \cite{ERV02,Esparza08}.

\begin{definition}[Causality, conflict, concurrency]
\label{def:causality}
Let $\net=\tup{P,T,F}$ be a net and $x,y\in P\cup T$ two nodes
of $\net$.
We say that
$x$ is a \emph{causal predecessor} of~$y$, noted $x<y$, if there exists a
non-empty path of arcs from $x$ to $y$. We note $x\leq y$ if $x<y$ or $x=y$.
If $x\leq y$ or $y\leq x$, then $x$ and $y$ are said to be
\emph{causally related}.
Transitions $u$ and $v$ are in \emph{direct conflict}, noted $u\dircf v$, iff $\preset{u}\cap\preset{v}\ne
\emptyset$; nodes
$x$ and $y$ are \emph{in conflict}, noted $x\conflict y$, if there exist
$u,v\in T$ such that $u\ne v$, $u\le x$, $v\le y$, and $u\dircf v$. We call
$x$ and $y$ \emph{concurrent}, noted $x\concurrent y$, if they are neither
causally related nor in conflict.
A set of concurrent places is called a \emph{co-set}.
\end{definition}

\begin{definition}[Occurrence net]
\label{def:occurrencenet}{\it
Let $\onet=\tup{\oplace,\otrans,\oflow,\cutn_0}$ be a Petri net. We say that $\onet$
is an \emph{occurrence net} if it satisfies the following properties:
\begin{enumerate}
\item The causality relation $<$ is acyclic;
\item $|\preset{\oplacen}|\le 1$ for all places $\oplacen\in\oplace$,
  and $\oplacen\in \cutn_0$ iff $|\preset{\oplacen}|=0$;
\item For every transition $\otransn\in\otrans$, $\otransn\conflict\otransn$ does not hold, and \(\{x \mid x \leq \otransn\}\) is finite.
\end{enumerate}}
\end{definition}

Following the  convention in the unfolding literature, we refer to the
places of an occurrence net as \emph{conditions} and to its transitions
as \emph{events}. Due to the structural constraints, the firing
sequences of occurrence nets have special properties: if some condition $\oplacen$ is
marked during a run, then the token on $\oplacen$ was either present initially
or produced by one particular event (the single event in~$\preset{\oplacen}$);
moreover, once the token on $\oplacen$ is consumed, it can never be replaced by
another token, due to acyclicity of~$<$. 
\begin{definition}[Configurations, cuts]
\label{def:configuration}{\it
Let $\onet=\tup{\oplace,\otrans,\oflow,\cutn_0}$ be an occurrence net.
A set $\confign\subseteq \otrans$ is called a
\emph{configuration} of $\onet$ if (i) $\confign$ is \emph{causally
closed}, i.e.\ $e'<e$ and $e\in \confign$ imply $e'\in \confign$; and (ii) $\confign$ is \emph{conflict-free}, i.e.\ if $e,e'\in\confign$,
then $\neg(e\conflict e')$. 
In particular, for any $e\in E$, $\cone{e}\define\{e'\in E:~e'\myleq e\}$ and $\trunk{e}\define\{e'\in E:~e'< e\}$ are configurations, called the \emph{cone} and \emph{stump} of $\otransn$, respectively; any $\confign$ such that $\exists~e\in E:~\confign=\cone{e}$ is called a \emph{prime configuration}.
Denote the set of all configurations of $\onet$ as $\configs(\onet)$, and its subset containing all \textbf{finite} configurations as $\finfigs(\onet)$, where we drop the reference to $\onet$ if no confusion can arise. The \emph{cut} of a finite $\confign$, denoted $\cut(\confign)$, is the set of
conditions $(\cutn_0\cup\postset{\confign})\setminus\preset{\confign}$. 
A \emph{run} is a maximal element of $\configs(\onet)$ w.r.t.\ set inclusion; denote the set of $\onet$'s runs as $\Omega=\Omega(\onet)$, and its elements generically by $\omega$.
If $\confign\in\finfigs$, let the \emph{crest} of $\confign$ be the set 
$\crest{\confign}\define \max_<(\confign)$ of its maximal events. 
We say that configuration $\confign$ \emph{enables} event $\otransn$, written $ \confign\omove{\otransn}$, iff i) $\otransn\not\in\confign$ and ii) $\confign\cup\{\otransn\}$ is a configuration.
} 
Configurations $\confign_1,\confign_2$ are in conflict, written $\confign_1\conflict\confign_2$, iff $(\confign_1\cup\confign_2)\not\in\configs$ or, equivalently, iff there exist $\otransn_1\in\confign_1$ and $\otransn_2\in\confign_2$ such that $\otransn_1\conflict\otransn_2$.\footnote{The use of the same symbol $\conflict$ is motivated by the fact that $\confign_1=\cone{\otransn_1}$ and $\confign_2=\cone{\otransn_2}$ implies $\confign_1\conflict\confign_2\Leftrightarrow \otransn_1\conflict\otransn_2$.}
\end{definition}

Intuitively, a configuration is a set of events that can fire during
a firing sequence of $\petrinet$, and its cut is the set of conditions
marked after that firing sequence. Note that $\emptyset$ is a configuration, that $\crest{\emptyset}=\emptyset$,
and that $\cutn_0$ is the cut of the configuration $\emptyset$. The crest of a prime configuration $\cone{e}$ is $\{e\}$.
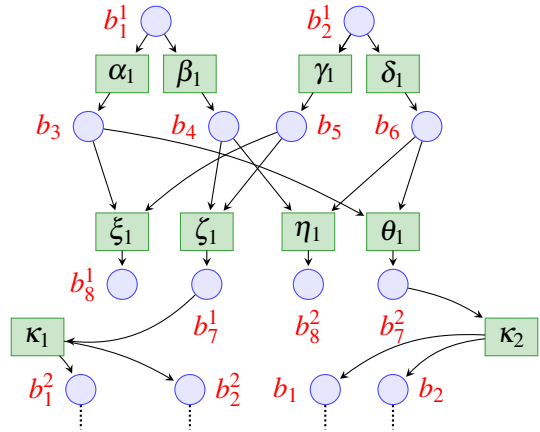
\begin{figure}[ht]
    \centering
\def\a{1}
\def\b{0.45}
\def\c{0.7}
\begin{tikzpicture}[>=stealth,shorten >=1pt,node distance=\a cm,auto]
  \node[cond] (b1) at (0, 0)  [label=left:\text{$b_1^1$}] {};
  \node[cond] (b2) at (6*\b, 0)  [label=left:\text{$b_2^1$}] {};
  \node[cond] (b1') at (5*\b, -7*\c)  [label=left:\text{$b_1$}] {};
  \node[cond] (b2') at (7*\b, -7*\c)  [label=right:\text{$b_2$}] {};
  \node[cond] (b1'') at (-1, -7*\c)  [label=left:\text{$b_1^2$}] {};
  \node[cond] (b2'') at (1*\b, -7*\c)  [label=right:\text{$b_2^2$}] {};
  \node[cond] (b3) at (-2*\b, -2*\c)  [label=left:\text{$b_3$}] {};
  \node[cond] (b4) at (2*\b, -2*\c)  [label=left:\text{$b_4$}] {};
  \node[cond] (b5) at (4*\b, -2*\c)  [label=right:\text{$b_5$}] {};
  \node[cond] (b6) at (8*\b, -2*\c)  [label=left:\text{$b_6$}] {};
  \node[cond] (b7a) at (1.5*\b, -5*\c)  [label=below:\text{$b_7^1$}] {};
  \node[cond] (b7b) at (7*\b, -5*\c)  [label=below:\text{$b_7^2$}] {};
  \node[cond] (b8a) at (-1*\b, -5*\c)  [label=left:\text{$b_8^1$}] {};
 \node[cond] (b8b) at (4.5*\b, -5*\c)  [label=below:\text{$b_8^2$}] {};

  \node[event] (a) at (-1*\b, -\c) {$\alpha_1$};
  \node[event] (b) at (1*\b, -\c) {$\beta_1$};

  \node[event] (x) at (-1*\b, -4*\c) {$\xi_1$};
\node[event] (c) at (5*\b, -\c) {$\gamma_1$};
\node[event] (d) at (7*\b, -\c) {$\delta_1$};
\node[event] (e) at (7*\b, -4*\c) {$\theta_1$};
\node[event] (r) at (10.5*\b, -6*\c) {$\kappa_2$};
\node[event] (rbis) at (-3.5*\b, -6*\c) {$\kappa_1$};
\node[event] (g) at (1.5*\b, -4*\c) {$\zeta_1$};
\node[event] (f) at (4.5*\b, -4*\c) {$\eta_1$};
\node (phantom1) at (-1, -8*\c)  [] {};
\node (phantom2) at (1*\b, -8*\c)  [] {};
\node (phantom3) at (5*\b, -8*\c)  [] {};
\node (phantom4) at (7*\b, -8*\c)  [] {};
 \path[->] (b1)  edge (a)
        (b3)  edge (x)
        (b3)  edge [bend left=10]    (e)
         (b1)  edge (b)
         (b2) edge (c)
         (b2) edge (d)
         (c) edge (b5)
        (b5) edge (g)
        (b5) edge [bend right=10] (x)
         (b4) edge (g)
         (b4) edge (f)
        (b6) edge (f)
          ; 
 \path[->] (b7b) edge [bend left=10] (r)
           (r) edge [bend right=20] (b2') 
           (r) edge [bend right=20] (b1')
 (b7a) edge [bend left=25] (rbis)
           (rbis) edge [bend left=10] (b2'') 
           (rbis) edge [bend left=10] (b1'')
 ;
  \path[->]  (a) edge (b3)
  (x) edge (b8a)
   (b) edge (b4)
    (d) edge (b6)
    (b2) edge (c)
(b6) edge  (e)
 (e) edge (b7b)
 (f) edge (b8b)
  (g) edge (b7a)  
 ;
 \draw[densely dotted, thick] (b1') -- (phantom3);
 \draw[densely dotted, thick] (b2') -- (phantom4);
 \draw[densely dotted, thick] (b1'') -- (phantom1);
 \draw[densely dotted, thick] (b2'') -- (phantom2);
\end{tikzpicture}
\caption{A prefix of the unfolding for the Petri net of Figure~\ref{fig:basinexample}.
    \label{fig:basinunfold}}
\end{figure}
In Figure~\ref{fig:basinunfold}, the initial cut is $\cutn_0=\{b_1^1,b_2^1\}$;
we have prime configurations, e.g., $\{\alpha_1\}$, $\{\beta_1\}$, $\{\xi_1\}$, $\{\zeta_1\}$ etc, 
and non-prime configurations $\{\alpha_1,\gamma_1\}$, $\{\alpha_1,\delta_1\}$ etc.

\textbf{Definition of Unfoldings.} Let $\petrinet=\tup{\place,\trans,\flow,\markn_0}$
be a safe Petri net. The unfolding $\unfolding=\tup{\oplace,\otrans,\oflow,\cutn_0}$ of
$\petrinet$ is an
 occurrence net (equipped with a mapping $\fold$)
 such that
the firing sequences and reachable markings of $\unfolding$ are exactly
the firing sequences and reachable markings of $\petrinet$ (modulo $\fold$), see below.
$\unfolding$ may be infinite; it can be inductively constructed as follows:

\begin{enumerate}
\item The condition set $\oplace$ is a subset of $(\otrans\cup\{\bot\})\times \place$.
 For a condition $\oplacen=\tup{\otransn,\placen}$, we will have $\otransn=\bot$ iff $\oplacen\in \cutn_0$;
 otherwise $\otransn$ is the singleton event in $\preset{\oplacen}$. Moreover, $\fold(\oplacen)=\placen$.
 The initial cut $\cutn_0$ contains exactly one condition $\tup{\bot,\placen}$
 for each initially marked place $\placen\in\markn_0$ of $\petrinet$.
\item The events of $\otrans$ are a subset of $2^\oplace\times \trans$. 
More precisely, for every co-set $\oplace'\subseteq\oplace$ such that
$\fold(\oplace')=
\preset{\transn}$,
we have an event $\otransn=\tup{\oplace',t}$.
 In this case, we add edges $\tup{\oplacen,\otransn}$
 for each $\oplacen\in \oplace'$ (i.e. $\preset{\otransn}=\oplace'$), we set $\fold(\otransn)=\transn$, and for each
 $\placen\in\postset{\transn}$, we add to $\oplace$ a condition $\oplacen=\tup{\otransn,\placen}$ connected by
 an edge $\tup{\otransn,\oplacen}$.
\end{enumerate}
Intuitively, a condition $\tup{\otransn,\placen}$ represents the possibility of putting
a token onto place $\placen$ through a particular set of events, while an event
$\tup{\oplace',\otransn}$ represents a possibility of firing transition $\otransn$ in a particular
context.

\textbf{Configurations and Markings.}
The following fact from the literature  will be used below:
\begin{lemma}[see e.g.~\cite{ERV02}]
\label{le:safe}{\it Fix a safe Petri net  $\petrinet=\tup{\place,\trans,\flow,\markn_0}$  and its unfolding $\unfolding=\tup{\oplace,\otrans,\oflow,\cutn_0, \fold}$.
 Then for any two conditions (events) $\oplacen,\oplacen'$ ($\otransn,\otransn'$) such that $\oplacen\concurrent\oplacen'$ ($\otransn \concurrent \otransn'$), one has $\fold(\oplacen)\ne\fold(\oplacen')$ ($\fold(\otransn)\ne\fold(\otransn')$). 
Moreover, every
 finite configuration $\confign$ of $\unfolding$ represents a possible firing
sequence whose resulting marking corresponds, due to the construction of
$\unfolding$, to a reachable marking of $\petrinet$. This marking is defined as
$\marking{\confign}\define\{\,\fold(\oplacen)\mid \oplacen\in\cut(\confign)\,\}$.}
\end{lemma}
This means, informally speaking, that any configuration of the system can be split into consecutive parts in such a way that each part is itself a configuration obtained by unfolding the Petri net `renewed' with the marking reached by the previous configuration. The following definition formalizes this.
\begin{definition}
Let $\onet=\tup{\oplace,\otrans,\oflow,\cutn_0}$ be an occurrence net.
For any finite configuration $\confign\in\finfigs(\onet)$, denote by $\onet_\confign\define\unfolding(\tup{\net,\marking{\confign}})$ the \emph{shift} of $\onet$ by $\confign$.
$\confign$ is the \emph{concatenation} of $\confign_1$ and $\confign_2$, written $\confign=\confign_1\concat\confign_2$,  iff one has 
\begin{enumerate}
    \item $\confign_1\in\finfigs(\onet)$ and
 $\confign_1\subseteq\confign$, 
    \item $\confign_2\in\finfigs(\onet_{\confign_1})$ and
  $\confign_2=\confign\backslash\confign_1$.
\end{enumerate}
\end{definition}
Clearly, the empty configuration $\emptyset$ satisfies $\confign\concat\emptyset=\emptyset\concat\confign=\confign$. If $\confign =\confign_1\concat\confign_2$, write $\confign_1=\confign\leftchop\confign_2$
and $\confign_2=\confign\rightchop\confign_1$. Moreover, write 
\begin{eqnarray*}
\confign=\konkat_{i=1}^n\confign_i&\text{iff}&
\confign=\confign_1\concat\ldots\concat\confign_n.
\end{eqnarray*}

In figure~\ref{fig:basinunfold}, setting $\confign_1\define\{\beta_1,\gamma_1\}$ , $\confign_2\define\{\zeta_1,\kappa_1\}$ and $\confign_3\define\{\beta_1,\gamma_1,\zeta_1,\kappa_1\}$, one has $\confign_3=\confign_1\concat\confign_2$ and consequently $\confign_1=\confign_3\leftchop\confign_2$ and $\confign_2=\confign_3\rightchop\confign_1$.

\textbf{Complete Prefix.}
In general, $\unfolding$
is an infinite net, but if $\petrinet$ is safe, then it is possible
to compute a finite prefix $\prefn$
of $\unfolding$ that is ``complete''
in the sense that every reachable marking of $\petrinet$ has a reachable
counterpart in $\prefn$,  and vice 
versa. 
\begin{definition}[complete prefix, see \cite{McM92,ERV02,Esparza08}]
\label{def:complete}
Let $\petrinet=\tup{\net,\markn_0}$ be a safe Petri net and
$\unfolding=\tup{\oplace,\otrans,\oflow,\cutn_0}$
its unfolding. A finite occurrence net ${\prefn}=\tup{\oplace',\otrans',\oflow',\cutn_0}$
is said to be a \emph{prefix} of $\unfolding$ if $\otrans'\subseteq \otrans$ is
causally closed, $\oplace'=\cutn_0\cup\postset{\otrans'}$, and $\oflow'$ is the restriction
of $\oflow$ to $\oplace'$ and $\otrans'$. A prefix $\prefn$ is said to be \emph{complete}
if for every reachable marking $\markn$ of $\petrinet$ there exists a
configuration $\confign$ of $\prefn$ such that (i) $\marking{\confign}=\markn$,
and (ii) for each transition $\transn\in \trans$ enabled in $\markn$, there is an event
$\tup{\oplace'',\transn}\in \otrans'$ enabled in $\cut(\confign)$. 
\end{definition}

We shall write $\prefixn_0=\prefixn_0(\petrinet)$ to denote an arbitrary complete prefix of the unfolding of $\netn$.
It is known (\cite{McM92,ERV02}) that the construction of such a complete prefix
is indeed possible, and efficient tools such as \textsc{Mole} (\cite{mole}) exist for this purpose.
While the precise details of this construction are out of scope for this paper;  some ingredients of it will play a role below, so we sketch them here.

\textbf{Complete prefix scheme.} The unfolding is stopped on each branch when some \emph{cutoff event} is added. The criterion for classifying an event $\otransn$ as cutoff is given by  Marking equivalence: the marking $\marking{\cone{\otransn}}$ that $\otransn$ `discovers' has already been discovered by a \emph{smaller} configuration. Now, the ordering relation $\adequate$ to compare two configurations must be an \emph{adequate} order, i.e. $\confign_1\subseteq\confign_2$ must imply $\confign_1\adequate\confign_2$, to ensure the completeness of the prefix obtained. As shown in~\cite{ERV02}, for some  choices of $\prec$, the obtained prefix may be bigger than the reachability graph for some safe nets; however, if $\prec$ is a \emph{total} order, the number of non-cutoff events of the prefix $\prefixn_0$ thus obtained never exceeds the size of the reachability graph.

We will assume throughout this paper that complete prefixes are computed according to some adequate total order, as is done in particular in the \textsc{Mole} tool~\cite{mole}. Below, we will propose a new such order relation that underlies a novel concept of distance between markings. 

\textbf{The nested family $(\prefixn_n)_{n\geq 0}$ of finite prefixes.} Denote the  complete prefix for $\petrinet$ obtained according to definition~\ref{def:complete} as $\prefixn_0$; we extend $\prefixn_0$ to increasing prefixes $\prefixn_1,\prefixn_1, \ldots$ as follows.
  Starting at $n=0$,
\begin{itemize}
    \item let $\configs^n\define\max(\configs(\prefixn_n))$,
\item set $
\markis_n\define \{\markn\in 2^\place:~\exists~\confign\in\configs^n:~\markn=\marking{\confign}\}$,
\item 
for all~$\markn\in\markis_n$, compute a complete prefix $\prefix^\markn$ of $\tup{\net,\markn}$; 
\item obtain $\prefixn_{n+1}$ by appending, to every $\confign\in\configs^n$, a copy of $\prefix^{\marking{\confign}}$ to every $\confign\in\configs^n$. 
\end{itemize}


\section{Doomed configurations, and how to avoid them}
\subsection{Bad, Free and Doomed Configurations and Markings.}
In this section, we present an algorithm that identifies precisely those configurations of a Petri net unfolding from which one can no longer avoid reaching a certain long-term behaviour, its theoretical foundations, and some experimental results. 
The formal setting here contains and extends the one established in~\cite{GiuaXie2005}, specialized to the 1-safe case. We assume that we are given a set of \emph{bad markings} $\badstates\subseteq 2^{\places}$. Since we are interested in long-term behaviours, we assume that $\badstates$ is reachability-closed, i.e. $\markn\in\badstates$ and $\markn\move{}\markn'$ imply $\markn'\in\badstates$. 

Define $\badconfs\define\{\confign\in\finfigs:\marking{\confign}\in\badstates\}$ as the set of \emph{bad configurations}, and let $\badconfs_0$ be the set of configurations in $\badconfs$ that are contained in $\prefixn_0$. 
$\badconfs\subseteq\configs$  is \emph{absorbing} or \emph{upward closed}, that is,  
   for all $\confign_1\in\badconfs$ and $\confign_2\in\finfigs$ such that $\confign_1\subseteq\confign_2$,  one must have $\confign_2\in\badconfs$.

For any  $\confign\in\configs$, let $\runs_\confign\define \left\{ \runn \in\runs:~\confign\subseteq \runn\right\}$ denote the maximal runs into which $\confign$ can evolve.
We are interested in 
those finite configurations all of whose maximal extensions are `bad', where we consider infinite configurations as bad if they contain a bad finite configuration. We will call such configurations \emph{doomed}, since from them, the system cannot avoid entering a bad marking sooner or later (and from then on, all reachable markings are bad).
\begin{definition}{\it
Configuration $\confign\in\finfigs$ is \emph{doomed} iff
\begin{eqnarray}
\label{eq:latbad} \forall~\runn\in\runs_\confign:~\exists~\confign'\in\finfigs:
\left\{\begin{array}{l}
\confign\subseteq\confign'\subseteq\runn\\
\land~\marking{\confign'}\in\badstates
\end{array}
\right.
\end{eqnarray}
The set of doomed configurations is denoted $\doofigs$; denote the set of minimal elements in $\doofigs$ by $\mindofigs$.
If $\confign$ is not doomed, it has at least one maximal extension that never reaches bad markings. We  call configurations that are not doomed \emph{free}, and denote the set of free configurations by $\viables$.}
\end{definition}

All reachable markings are represented by at least one configuration.
Moreover, since the future evolution of $\netn$ depends only on the current marking,  
$\marking{\confign_1}=\marking{\confign_2}$ for two configurations $\confign_1$ and $\confign_2$ implies that either both $\confign_1$ and $\confign_2$ are free, or both are doomed. 
Therefore, by extension, we call $\marking{\confign}$ free or doomed whenever $\confign$ is.

\textit{Running Example.} In the context of Figures~\ref{fig:basinexample} and~\ref{fig:basinunfold}, we consider $\badstates$ the singleton set containing the marking $\markn_8=\{\place_8\}$. Clearly, $\confign_1\define\{\alpha_1,\gamma_1,\xi_1\}$ and $\confign_2=\{\beta_1,\delta_1,\eta_1\}$ satisfy $\marking{\confign_1}=\marking{\confign_2}=\markn_8$ and therefore $\confign_1,\confign_2\in\badconfs$. But note that $\confign_1'\define\{\alpha_1,\gamma_1\}$ and $\confign_2'=\{\beta_1,\delta_1\}$ produce markings outside  $\badstates$, but they are doomed since any extension of these configurations leads into $\badstates$. Therefore, $\confign_1',\confign_2'\in\badconfs$. On the other hand,
$\emptyset$ is free, as well as $\{\beta_1,\gamma_1\}$, $\{\alpha_1,\delta_1\}$, etc.
We note in passing that the Petri net in Fig~\ref{fig:basinexample} allows to refine the understanding of the `tipping point' by showing that doom is not brought about by a single transition but rather the combined effect of two independent choices; this fact is obscured, or at least far from obvious, in the state graph shown in Figure~\ref{fig:basinstategr}.

Identifying free and doomed configurations belongs to the core objectives of this paper. In a first step towards that, Theorem~\ref{th:notbad}  below uses a similar proof idea as Lemma 8 in~\cite{HRS-acsd13} in the context of fault  diagnosis.
Let us first recall the notion of \emph{spoilers}, introduced in  ~\cite{HRS-acsd13}:
\begin{definition}\textit{
A \emph{spoiler} of transition $\transn$ (or  event $\otransn$) is any $\transn'\in\trans$ ($\otransn'\in\otrans$) such that $\preset{\transn'}\cap\preset{\transn}\ne\emptyset$ ($\preset{\otransn}\cap\preset{\otransn'}\ne\emptyset$).
We write $\spoil{\transn}$  ($\spoil{\otransn}$) for the set of $\transn'$s ($\otransn$'s) spoilers.}
\end{definition}
Note that $\transn\in\spoil\transn$ for all $\transn\in\trans$. The spoilers of $\transn$ are characterized by the fact that their firing cancels any enabling of $\transn$; that is, by being either in conflict with $\transn$, or identical with $\transn$.
\begin{theorem}\label{th:notbad}{\it
A configuration $\confign\in\finfigs$ is \textbf{free} iff either a) there exists a finite maximal configuration $\confign'$ such that $\confign\subseteq\confign'\not\in\badconfs$, or b) there exist  configurations $\confign_1,\confign_2\in\finfigs$ such that 
\begin{enumerate}
    \item\label{con:nest} $\confign\subseteq\confign_1\subseteq \confign_2\notin\badconfs$;
    \item \label{con:mark}$\marking{\confign_1}=\marking{\confign_2}$;
    \item \label{con:spoil} for all events $\otransn\in\otrans$ such that $\confign_1\omove{\otransn}$, one has $\spoil{\otransn}\cap \confign_2\ne\emptyset$.
\end{enumerate}
}
\end{theorem}
Some comments are in order before giving the proof of Theorem~\ref{th:notbad}.
First of all, the requirement to check whether $\confign_2\not\in\badconfs$ can be met by checking whether $\marking{\confign_2}\in\clbadstates$.
Second, the spoiling condition (3) ensures that the process that takes $\confign_1$ to $\confign_2$ forms a loop whose iteration yields a run.

\bepr In the following, let $M:=\marking{C_1}$. We first prove the right-to-left implication. Case a) is obvious, so assume that b) holds.
Let $\confign_2=\confign_1\concat\hat{\confign}$; then by 2., we can append $\hat{\confign}$ to $\confign_2$, yielding a strictly increasing sequence of configurations $(\confign_n)_{n\in\NN}$
such that $\confign_{n+1}=\confign_n\concat \hat{\confign}$, and $\marking{\confign_n}=\markn$ for all $n$. By property~3, we know that $\hat{\confign}$ contains spoilers for all its initially enabled events , hence 
no transition remains enabled forever, and $\omega\define\bigcup_{n\in\NN}\confign_n$ is a maximal configuration. It remains to show that $\omega$ contains no bad configuration:
Suppose that there is $C'\subseteq\omega$ with $C'\in\badconfs$. Since $C'$ is finite, we have $C'\subseteq C_n$ for some $n$. But then, $M=\marking{C_n}$ is reachable from $\marking{C'}\in\badstates$,
contradicting our assumptions. Thus $\omega$ never enters a bad state, and $\confign$ is free.

For the forward implication, assume that $\confign$ is free. Then there exists $\runn\in\runs_\confign$ such that $C'\notin\badconfs$ for all finite $C'\subseteq\runn$. If this $\runn$ can be chosen finite, then a) holds and we are done; so assume henceforth that $\runn$ must be chosen infinite. Clearly, there must exist a reachable marking $\markn$ that is visited an infinite number of times  by a family of  nested finite configurations $(\confign^n)_{n\in\NN}$ such that $\bigcup_{n\in\NN}\confign^n=\runn$. Let $\confign_1\define\confign^1$ and  $E':=\{\otransn:~\confign_1\omove{\otransn}\}$. Let $K$ be the smallest index such that for all $\otransn\in E'$, one has  $\spoil{\otransn}\cap\confign^K\ne\emptyset$; such a $K$ must exist since $\omega$ is maximal. Then $\confign_1$ and $\confign_2\define\confign^K$ have the required properties.
\eepr

Notice that the proof could be restructured by observing that case a) of Theorem~\ref{th:notbad} is indeed a special instance of case b). In fact, taking $\confign_1\define\confign_2\define\confign'$ with  $\confign'$ according to  case a), conditions 1 and 2 of part b) are obviously satisfied, and condition~3 holds vacuously since no event is enabled in $\confign_1=\confign'$.  We note in passeing that this observation is helpful in simplifying the implementation used for the experiments  below.

The interest of Theorem~\ref{th:notbad} lies in the following fact:
\begin{lemma}
\label{le:complix} For $\confign\in\finfigs$, checking whether $\confign$ is free can be done using finite prefix $\prefixn_1$ of $\unfolding(\net,\marking{\confign})$.
\end{lemma}
\bepr If $\confign$ is free, let $C_1$ and $C_2$ be the configurations witnessing this fact from Theorem~\ref{th:notbad}, and let $M:=\marking{C_1}=\marking{C_2}$. If such configurations exist, then $C_1$ can be chosen from the complete prefix $\prefixn_0$, and $C_2$ can be chosen from $\prefixn_1$, notably 
in the copy of  $\prefixn_{\marking{\confign_1}}$  appended after $\confign_1$. 

Checking whether the  configuration $\confign_2$ thus found is in $\badconfs$ is immediate, since it suffices to check whether its marking is in $\badstates$, using the fact that  $\badstates$ is reachability-closed. 
To check the spoiler condition (3) of Theorem~\ref{th:notbad}, it suffices to check whether the conditions of the cut of $\confign_1$ that are not consumed by $\confign_2$ enable some event.
\eepr

\subsection{Finding Minimally Doomed Configurations: Algorithm \textsc{Mindoo}}
\textbf{Shaving and Rubbing.} Let us start by observing that $\badconfs$, an upward closed set by construction, also has some downward closure properties, meaning one can restrict control to act on `small' configurations.
\begin{definition}
{\it An event $\otransn$ is \emph{unchallenged} iff there is no $\otransn'$ such that $\otransn\dircf\otransn'$, i.e. $\postset{(\preset{\otransn})}=\{\otransn\}$. 
}
\end{definition}
\begin{lemma}\label{le:shave}
Let $\confign\in\finfigs$ and $\otransn\in\crest{\confign}$ unchallenged; set $\confign'\define\confign\backslash\{\otransn\}$. Then $\confign'\in\finfigs$, and $\runs_\confign=\runs_{\confign'}$.
\end{lemma}
\bepr 
$\confign'\in\finfigs$ holds by construction. Also, $\runs_{\confign}\subseteq\runs_{\confign'}$ follows from $\confign'\subseteq\confign$; it remains to show the reverse inclusion. Assume there exists $\runn\in\runs_{\confign'}\backslash\runs_\confign$; then $\confign\backslash\runn=\{\otransn\}$, and $\trunk{\otransn}\subseteq\runn$. By maximality, $\runn$ must contain some $\otransn'$ such that $\otransn\conflict\otransn'$. Then by definition, there are events $u\ne v$, $u\le e$, $v\le e'$, and $u\dircf v$.
In particular, $u\conflict e'$, and since $\{e'\}\cup\trunk{\otransn}\subseteq\runn$,
this implies $u=e$. But $e$ is unchallenged, so $v$ cannot exist, and neither can $\runn$.

\eepr

\begin{figure}
    \centering
 \def\a{1}
\def\b{0.95}
\def\c{0.85}
\begin{tikzpicture}[>=stealth,shorten >=1pt,node distance=\a cm,auto]
  \node[cond] (b1) at (0, 0)  [label=left:\text{$b_1$}] {};
  \node[cond] (b2) at (-1*\b, -2*\c)  [label=left:\text{$b_2$}] {};
  \node[cond] (b3) at (1*\b, -2*\c)  [label=right:\text{$b_3$}] {};
  \node[cond] (b4) at (-1*\b, -4*\c)  [label=right:\text{$b_4$}] {};
  \node[cond] (b5) at (1*\b, -4*\c)  [label=left:\text{$b_5$}] {};
  \node[cond] (b6) at (-3*\b, -6*\c)  [label=left:\text{$b_6$}] {};
  \node[cond] (b7) at (-1*\b, -6*\c)  [label=left:\text{$b_7$}] {};
  \node[cond] (b8) at (1*\b, -6*\c)  [label=right:\text{$b_8$}] {};
  \node[cond] (b9) at (3*\b, -6*\c)  [label=right:\text{$b_9$}] {};
 \node[cond] (b10) at (0*\b, -8*\c)  [label=left:\text{$b_{10}$}] {};

  \node[event] (x) at (0*\b, -\c) {$x$};
  \node[event] (y) at (-1*\b, -3*\c) {$y$};
  \node[event] (z) at (1*\b, -3*\c) {$z$};
\node[event] (a) at (-3*\b, -5*\c) {$\alpha$};
\node[event] (b) at (-1*\b, -5*\c) {$\beta$};
\node[event] (c) at (1*\b, -5*\c) {$\gamma$};
\node[event] (d) at (3*\b, -5*\c) {$\delta$};
\node[event] (u) at (0*\b, -7*\c) {$u$};

 \path[->] (b1)  edge (x)
       (x)  edge (b2)
      (x)  edge (b3)
      (b2) edge (y)
      (b3) edge (z)
      (y) edge (b4)
      (z) edge (b5)
      (b4) edge (a)
      (b4) edge (b)
      (b5) edge (c)
      (b5) edge (d)
      (a) edge (b6)
      (b) edge (b7)
      (c) edge (b8)
      (d) edge (b9)
      (b7) edge (u)
      (b8) edge (u)
      (u) edge (b10)
       ;
\end{tikzpicture}
    \caption{An occurrence net. With $\confign\define\{x,y,z,\beta,\gamma\}$ and $\confign'\define\confign\cup\{u\}$, suppose $\badstates=\{\marking{\confign'}\}=\fold(\{b_{10}\})$. Then  $\shaved{\confign'}=
    \confign$, and $\confign$ is doomed. Moreover, $\confign\in\mindofigs$ since both 
    $ \confign_4\define\confign\backslash\{\beta\}$ and $ \confign_5\define\confign\backslash\{\gamma\}$ are free.}
    \label{fig:wreath}
\end{figure}
\begin{definition}
A configuration $\confign\in\finfigs$ such that $\crest{\confign}$ contains no unchallenged event is called \emph{shaved}. 
\end{definition}
Clearly, every $\confign\in\finfigs$ contains a unique maximal shaved configuration, which we call $\shaved{\confign}$; it can be obtained from $\confign$ by recursively `shaving away' any unchallenged $\otransn\in\crest{\confign}$, and then continuing with the new crest, until no unchallenged events remain.

\textit{Example.} In the context of Figure~\ref{fig:wreath}, for $\confign_1=\{x,y,z\}$ and $\confign_2=\confign_1\cup\{\beta,\gamma,u\}$, one has $\shaved{\confign_1}=\emptyset$ since $x$, $y$, and $z$ are unchallenged, and $\shaved{\confign_2}=\confign_1\cup\{\beta,\gamma\}$ since $u$ is unchallenged but neither $\beta$ nor $\gamma$ are. 
Note that in the unfolding of the running example shown in Figure~\ref{fig:basinunfold}, the $\kappa$-labeled events are the only unchallenged ones.

As a consequence of Lemma~\ref{le:shave}, any $\confign\in\finfigs$ is in $\badconfs$ iff $\shaved{\confign}$ is. Still,  it may be possible that such a $\shaved{\confign}$ can still be reduced further by removing some of its crest events. This would be the case, e.g., if two conflicting events both lead to a bad state. Thus, given a crest event $e$, we test whether $\confign\backslash\{\otransn\}$ is free (e.g. because some event in conflict with $\otransn$ may allow to move away from doom) or still doomed. If the latter is the case, then $\confign$ was not minimally doomed, and analysis continues with $\confign\backslash\{\otransn\}$ (we say that we `rub away' $\otransn$). If $\confign\backslash\{\otransn\}$ is free, we leave $\otransn$ in place and test the remaining events from $\crest{\confign}$. A configuration that is shaved and from which no event can be rubbed away is minimally doomed. 

Algorithm~\ref{alg:two} uses a `worklist' set $\worklist$ of doomed, shaved configurations to be explored; $\worklist$ is modified when a configuration is replaced by a set of rubbed (and again, shaved) versions of itself, or when a configuration $\confign$ is identified as minimally doomed, in which case it is removed from $\worklist$
 and added to $\mindooout$.
\begin{algorithm}[hbt!]
\caption{Algorithm \textsc{MinDoo}}\label{alg:two}
\KwData{Safe Petri Net $\petrinet=\tup{\place,\trans,\flow,\markn_0}$ and $\badstates\subseteq 2^{\place}$}
\KwResult{The set $\mindooout$ of $\petrinet's$ $\subseteq$-minimal doomed configurations}
$\mindooout\gets\emptyset$;~
$\worklist \gets \emptyset$\;
\ForEach{$\confign\in\min_{\subseteq}(\badconfs_0)$}{
    $\confign'\gets\shaved{\confign}$\;
    $\worklist\gets\worklist\cup\{\confign'\}$\;
}
\While{$\worklist\neq\emptyset$}{
    Pick $\confign\in\worklist$;~
    $\addit\gets \true$\;
    \uIf{$(\confign\backslash\crest{\confign})$ is doomed}{
            $\addit\gets \false$\;
            $\confign'\gets\shaved{\confign\backslash\crest{\confign}}$\;
            $\worklist \gets \worklist\cup\{\confign'\}$\;
            }
    \Else{
        \ForEach{$\otransn\in\crest{\confign}$}{
            \If{$(\confign\backslash\{\otransn\})$ is doomed}{
                $\addit\gets \false$\;
                $\confign'\gets\shaved{\confign\backslash\{\otransn\}}$\;
                $\worklist \gets \worklist\cup\{\confign'\}$\;
            }
    }}
    $\worklist\gets\worklist\backslash\{\confign\}$\;
    \If{$\addit$}{
        $\mindooout\gets\mindooout\cup\{\confign\}$\;
      }}
\Return{$\mindooout$}
\end{algorithm}

Every branch stops when a minimally doomed configuration is reached, i.e., a doomed configuration $\confign$ such by rubbing off any crest event $e$ from $\confign$ makes it free, i.e. $\confign\backslash\{e\}$ is free for all $e\in\crest{\confign}$. When the worklist is empty,  all minimally doomed configurations have been collected in~$\mindooout$.
Note that if $\emptyset\in\worklist$ at any stage during the execution of Algorithm~\textsc{Mindoo}, then
$\emptyset$ will be added to $\mindooout $, since \textsc{Mindoo} will not enter the second \textbf{foreach}-loop in that case.
In fact, if this situation arises, \emph{every} configuration is doomed, and thus  $\emptyset$ is the unique minimally doomed configuration.

The configurations produced in the course of the search strictly decrease w.r.t both size and inclusion. Moreover, an upper bound on the prefixes explored at each step is given by $\badconfs$, itself strictly contained in the complete finite prefix used to find all bad markings. 
According to~\cite{ERV02}, this prefix can be chosen of size equal or smaller (typically: considerably smaller) than the reachability graph of $\petrinet$.

\begin{theorem}{\it
For any safe Petri net $\petrinet=\tup{\net,\markn_0}$ and bad states set $\badstates\subseteq \reach_\net{(\markn_0)}$, Algorithm \textsc{MinDoo} terminates, with output set $\mindooout$ containing exactly all minimal doomed configurations, i.e. $\mindooout=\mindofigs$.}
\end{theorem}
\bepr 
Termination follows from the finiteness of $\min_\subseteq(\badconfs_0)$,
since in  each round of \textsc{MinDoo} there is one configuration $\confign$ that is either replaced by a set of strict prefixes or removed from $\worklist$. Therefore, after a finite number of steps $\worklist$ is
empty. 
According to Lemma~\ref{le:complix}, the status (doomed or free) of a given finite configuration can effectively be checked on a fixed finite prefix of $\unfolding$.
Assume that after termination of \textsc{MinDoo}, one has $\confign\in\mindooout$; we need to show $\confign\in\mindofigs$. Clearly, when $\confign$ was added to $\mindooout$, it had been detected as doomed; it remains to show that $\confign$ is also minimal with this property. Assume that there is $\confign'\subsetneq\confign$ that as doomed as well. But in that case  there exists $\otransn\in\crest{\confign}$  such that $\confign'\subseteq(\confign\backslash\{\otransn\})\subsetneq\confign$, which implies that this  $(\confign\backslash\{\otransn\})$ is doomed as well. But then $\addit$ has been set to $\false$ in the second \textbf{foreach}-loop, before $\confign$ could have been added to $\mindooout$. 

Conversely, let $\confign\in\mindofigs$. Then $(\confign\backslash\{\otransn\})$ is free for all $\otransn\in\crest{\confign}$; the variable $\addit$ remains thus at the value $\true$ because no round of the second \textbf{foreach}-loop can flip it. Thus $\confign$ is added to $\mindooout$, from which \textsc{MinDoo} never removes any configuration.
\eepr 

\subsection{Implementation and Experiments.}
A prototype implementation of \textsc{Mindoo} is available at~\cite{Mindoo}.
It takes as input a safe Petri net in the PEP format and relies on \textsc{Mole}~\cite{mole} for computing the
initial finite prefix $\prefixn_0$ and its extensions.
Algorithm~\ref{alg:two} is implemented in Python, where the identification of maximal
configurations, bad configurations, as well as the verification of doomed status of a configuration
is performed in Answer-Set Programming (ASP) employing the \textsc{Clingo} solver~\cite{clingo}, a logic programming technology close to SAT solving.

We illustrate in Table~\ref{tab:experiments} the behavior of the implementation on different
instances of Petri nets modeling biological processes.
\begin{table}
    \caption{Statistics of Algorithm~\ref{alg:two} on Petri net models of biological systems.
        The size of $\prefixn_0$ and $\prefixn_1$ is the number of their events;
        ``\# min doomed cfg'' is the number of minimally doomed configurations;
        ``\# doom checks'' is the number of SAT checks for doom status of a configuration.
    ``time'' is the total computation time on a 1.8Ghz CPU
    \label{tab:experiments}}
\begin{tabular}{|l|r|r|r|r|r|}\hline
Model & size $\prefixn_0$ & size $\prefixn_1$ & \# min doomed cfg & \# doom checks & time\\\hline
Lambda switch & 126 & 1,060 & 10 & 29 & 1s\\
Cell death receptor & 791 & 19,262 & 57 & 407 & 37s\\
Budding yeast cell cycle & 1,413 & 184,363 &  114 & 837 & 8m3s\\
\hline
\end{tabular}
\end{table}
In each case, we report the size (number of events) of prefixes $\prefixn_0$ and $\prefixn_1$
(including cut-off events), the number of minimally doomed configurations, and the number of
configurations which have been tested for being doomed.
The purpose of the conducted experiments was to study the tractability of our approach on literature models of biological systems for which the study of doomed configuration was relevant.
As exhibited in~\cite{CHJPS-cmsb14}, one of the first potential bottleneck is the tractability of the computation of the finite complete prefix $\prefixn_0$ and the enumeration of maximal configurations, which is required for computing $\prefixn_1$. Then, our experiments have focused on assessing how evolved the number of minimally doomed configurations, the number of candidate configurations screened by Algorithm~\ref{alg:two}, and the overall computation time, with different sizes of prefixes $\prefixn_1$.

We selected 3 models published as Boolean networks, which can be translated as safe Petri nets using the encoding described in \cite{CHJPS-cmsb14} implemented in the tool \textsc{Pint}~\cite{Pint-CMSB17}.
The ``Lambda switch'' model~\cite{tt95} comprises 11 places and 41 transitions, and possesses two limit
behaviors, one being a deadlock, marked as a bad marking.
The ``Cell death receptor'' model~\cite{calzone2010} comprises 22 places and 33 transitions, and reproduces a bifurcation
process into different cell fates, one of which has been declared as bad (apoptosis).
In these two cases, the minimally doomed configurations identify configurations in which a decisive event has just taken place, committing the system to the attractor marked as bad.
The ``Budding yeast cell cycle'' model~\cite{Orlando2008} comprises 18 places and 32 transitions, and represents the
oscillation of gene activity during the cell cycle. In this model, the cycle can exit and eventually
reach a marking corresponding to all genes being inactive, which is our bad marking.
In this later case, the minimally doomed configurations identify precisely when the system exits its oscillatory behavior.

It appears that the computation time for identifying minimally doomed configurations seems mostly affected by the size
of $\prefixn_1$ for the verification of the doom property of a configuration by ASP solving,
implementing the conditions of Theorem~\ref{th:notbad}.
In each case, the number of minimally doomed configurations is a fraction of the size of the finite complete prefix $\prefixn_0$. Future work may explore compact representations of the set of minimally doomed configurations, as they typically share a large amount of events, and may ease biological interpretations.

\section{Protectedness}

\subsection{Cliff-Edges and Ridges.} From the minimal doomed configurations, we derive the critical `points' at which a run becomes doomed:
\begin{definition}
An event set $\watershed\subseteq\otrans$ is called a \emph{cliff-edge} iff there exists a minimally doomed configuration $\confign\in\mindofigs$ such that $\watershed=\crest{\confign}$. The set of cliff-edges is denoted $\tips$. The folding $\ridge\define\fold(\watershed)\subseteq\trans$ of a cliff-edge $\watershed$ is called a \emph{ridge}.
\end{definition}
To complete the map of the evolutional landscape for $\netn$, it is important to find, in a bounded prefix of the unfolding, all ridges that determine the viability of a trajectory. Notice that the completeness of prefix $\prefixn_0$ only guarantees that all reachable \emph{markings} of $\netn$ are represented by at least one configuration of $\prefixn_0$; this does not extend to a guarantee that all concurrent steps that lead into a doomed marking can be found in $\prefixn_0$ as well. Fortunately, one has: 
\begin{lemma}
\label{le:ridge} For every ridge $\ridge$ of $\netn$ there is a witness in $\prefixn_0$, i.e. there exists a minimally doomed configuration $\confign$ in $\prefixn_1$ such that $\fold(\crest{\confign})=\ridge$.
\end{lemma}
\bepr
Fix $\ridge$, and let $\confign_\ridge$ be any configuration such that $\fold(\crest{\confign_\ridge})=\ridge$; set $\markn^\confign\define\marking{\confign_\ridge}$, and let $\markn_\ridge^\confign$ the unique reachable marking such that $\markn_\ridge^\confign\move{\ridge}\markn^\confign$. Then any such  $\markn_\ridge^\confign$ is represented by some $\confign^\ridge$ in $\prefixn_0$. By construction, there exists a cliff-edge $\watershed$ such that $\confign^\ridge\omove{\watershed}$ and $\fold(\watershed)=\ridge$. Then $\confign\define\confign^\ridge\cup\watershed$ is a minimally doomed configuration that lies within $\prefixn_1$. 
\eepr

\subsection{Measuring the Distance from Doom}
With the above, we have the tools to draw a map of the `landscape' in which the system evolves, with doomed zones and cliff-edges highlighted. What we wish to add now is to assist \emph{navigation} in this landscape: we intend to give a meaningful measure of how well, or badly, a current system state is protected against falling from a cliff-edge. We chose to measure this distance not in terms of the \emph{length} of paths, or similar notions, but rather in terms of the \emph{choices} that are made by the system in following  a particular path.

Consider a  configuration $\confign$ and the nonsequential process that it represents. Some of the events in $\confign$ can be seen as representing a \emph{decision}, in the sense that their occurrence took place in  conflict with some event that was enabled by some prefix of  $\confign$. The number of such events gives a measure of the information contained in $\confign$, in terms of the decisions necessary to obtain $\confign$:
\begin{definition}
Let $\confign\in\finfigs$, and define
\begin{eqnarray*}\label{eq:decision}
\dheight{\confign}&\define&\left|\left\{\otransn\in\confign:~\exists~\otransn'\in\otrans:~
\otransn\strcf^\confign\otransn'
\right\}\right|,
\end{eqnarray*}
where $\strcf$ is the \emph{strict $\confign$-conflict} relation defined, for all $\otransn\in\confign$, by
\begin{eqnarray*}
\otransn\strcf^\confign\otransn'&\setgdw&\otransn\dircf\otransn'~\land~\trunk{\otransn'}\subseteq\confign.
\end{eqnarray*}

$\dheight{\confign}$ is  called the \emph{decisional height} of $\confign$.
\end{definition}
In Figure~\ref{fig:basinunfold}, the configuration $\confign_1=\{\xi_1,\alpha_1,\gamma_1\}$  satisfies $\dheight{\confign_1}=2$, whereas for $\confign_0=\{\beta_1\}$, one has $\dheight{\confign_0}=1$.

Note that $\strcf^\confign$ is more restrictive than direct conflict $\dircf$; it is also more restrictive than  the 
\emph{immediate conflict} in the literature (e.g.~\cite{abbes:hal-00350226}). It is closely dependent on the configuration $\confign$ under study, and describes precisely those events \emph{against} which the process had to decide in performing $\confign$.

\begin{figure}
    \centering
    \def\a{1}
\def\b{0.95}
\def\c{0.85}
\begin{tikzpicture}[>=stealth,shorten >=1pt,node distance=\a cm,auto]
  \node[cond] (b1) at (-1*\b0, 0)  [label=above:\text{$b_1$}] {};
  \node[cond] (b2) at (1*\b,0)  [label=above:\text{$b_2$}] {};
  \node[cond] (b3) at (-\b, -2*\c)  [label=left:\text{$b_3$}] {};
  \node[cond] (b4) at (1*\b, -2*\c)  [label=left:\text{$b_4$}] {};
  \node[cond] (b5) at (3*\b, -2*\c)  [label=right:\text{$b_5$}] {};
  \node[cond] (b6) at (-2*\b, -4*\c)  [label=left:\text{$b_6$}] {};
  \node[cond] (b7) at (0*\b, -4*\c)  [label=left:\text{$b_7$}] {};
  \node[cond] (b8) at (2*\b, -4*\c)  [label=right:\text{$b_8$}] {};

  \node[event] (x) at (-\b, -\c) {$x$};
  \node[event] (y) at (1*\b, -\c) {$y$};
  \node[event] (z) at (3*\b, -\c) {$z$};
\node[event] (a) at (-2*\b, -3*\c) {$\alpha$};
\node[event] (b) at (-0*\b, -3*\c) {$\beta$};
\node[event] (c) at (2*\b, -3*\c) {$\gamma$};

 \path[->] (b1)  edge (x)
        (x)  edge (b3)
      (b2) edge (y)
      (b2) edge (z)
       (y) edge (b4)
      (z) edge (b5)
      (b3) edge (a)
   (b3) edge (b)
      (b4) edge (b)
      (b4) edge (c)
      (a) edge (b6)
      (b) edge (b7)
      (c) edge (b8)
       ;
\end{tikzpicture}
    \caption{Illustration of direct conflict.}
    \label{fig:onconflicts}
\end{figure}
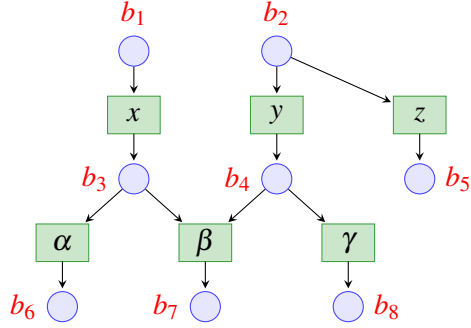

Now, for any free marking $\markn$ (or, equivalently, any free configuration $\confign$ such that $\marking{\confign}=\markn$),
we wish to measure the threat represented by doomed markings reachable from $\markn$: how far away from doom is the system when it is in $\markn$ ? Using the decisional height introduced above, we can define a height difference in terms of the conflicts that lead from one marking to another: 
\begin{definition}
For $\confign\in\finfigs$, let 
\begin{eqnarray}
\mindofigs_\confign&\define & \left\{ \begin{array}{lcr}
\{\confign'\in\mindofigs:~\confign\subseteq\confign'\}&: & 
\confign\in\viables\\
    \{\confign\} &:& \confign\in\badconfs
\end{array}\right.
\end{eqnarray}
The \emph{protectedness} of $\confign$ is then
\begin{eqnarray}
\schutz{\confign}&\define& \min_{\confign'\in\mindofigs_\confign}\left\{\dheight{\confign'\rightchop\confign}\right\}
 \end{eqnarray}
 
\end{definition}
 In Figure~\ref{fig:wreath}, with the definitions introduced there, $\schutz{\confign}=\schutz{\confign'}=0$. Setting $\confign_1\define\{x\}$, $\confign_2\define\{x,y\}$, $\confign_3\define\{x,z\}$, $\confign_4\define\confign_2\cup\confign_3$, $\confign_5\define\confign_4\cup\{\beta\}$,
and $\confign_6\define\confign_4\cup\{\gamma\}$,
one further has 
\begin{eqnarray*}
\schutz{\confign_1}=\schutz{\confign_2}=\schutz{\confign_3}&=&2\\
\schutz{\confign_4}=\schutz{\confign_5}&=&1.
\end{eqnarray*}
Returning to Figure~\ref{fig:onconflicts}, suppose that $\confign'=\{x,y,\beta\}$ is the only minimally doomed configuration. Then for $\confign=\{x,z,\alpha\}$ as above, we have $\schutz{\confign}=1$, because the only direct conflict here is the one between $z$ and $y$. 

Note that the definition of protectedness is parametrized by the choice of conflict relation in computing $\dheight{\bullet}$. Using direct conflict instead of strict conflict would increase $\dheight{\bullet}$ and lead to an overevaluation of protectedness.

To see the point, consider the occurrence net in Figure~\ref{fig:onconflicts}. Let $\confign_\alpha=\{x,z,\alpha\}$, $\confign_\beta=\{x,y,\beta\}$  and $\confign_\gamma=\{x,y,\alpha,\gamma\}$. We have
$\dheight{\confign_\alpha}=1$, $\dheight{\confign_\beta}=3$ and $\dheight{\confign_\gamma}=2$. Were $\strcf$ replaced by $\dircf$ in the computation of $\dheight{\bullet}$, these values would not change \emph{except} for $\confign_\alpha$ where it would change to $2$. As a result, if $\confign\in\mindofigs$, the protectedness of the empty configuration would be evaluated as $2$, whereas by our definition $\schutz{\emptyset}=1$. Indeed, $\emptyset$ is \emph{just one wrong decision away from doom}, and this is what protectness is meant to express.

\subsection{Computing Protectedness is Feasible}
Computation of $\schutz{\bullet}$ does not require any larger data structure than those already required for  computing $\mindofigs$ according to Lemma~\ref{le:complix}:
\begin{lemma}
\label{fig:schutz} There is a complete prefix scheme  producing a complete prefix $\prefixn_0$  whose size is bounded by the number of reachable markings, and such that for every finite configuration $\confign,$ $\schutz{\confign}$ can be computed on $\prefixn_0(\marking{\confign})$.
\end{lemma}
\bepr If $\mindofigs\cap\configs(\prefixn_0)=\emptyset$, then all extensions of $\confign$ are free, and we are done. Otherwise,
the crucial step is to find an adequate \emph{total} order $\adequate$ on finite configurations, that ensures that $\prefixn_0$ contains at least one  minimally doomed configuration that  minimizes $\dheight{\bullet}$ over all minimally doomed configurations in  $\unfolding(\marking{\confign})$.
The following order $\prec$ is obtained by modifying the total order $\prec_F$ introduced in~\cite{ERV02}, Def. 6.2.:
For $\confign_1,\confign_2\in\finfigs$, write $\confign_1\adequate\confign_2$ iff either 
\begin{itemize}
    \item $\dheight{\confign_1}<\dheight{\confign_2}$, or
    \item $\dheight{\confign_1}=\dheight{\confign_2}$ and $\confign_1\ll\confign_2$, or
    \item $\dheight{\confign_1}=\dheight{\confign_2}$ and $\confign_1\equiv\confign_2$, and $FC(\confign_1)\ll FC(\confign_2)$,
\end{itemize}
where $\ll$ ($\equiv$) denote lexicographic ordering (lexicographic equivalence) wrt some total ordering of the transition set $\trans$, and $FC$ denotes Cartier-Foata normal form. The proof of Theorem 6.4. of~\cite{ERV02} extends immediately, proving that $\prec$ is an adequate total order; therefore, Lemma 5.3. of~\cite{ERV02} applies, hence any complete prefix $\prefixn_0^\prec$ obtained via the scheme using $\prec$ is bounded in size by the reachability graph. Now, let $\configs^*$ be the set of configurations from $\mindofigs(\marking{\confign})$ that minimize $\dheight{\bullet}$; by construction of $\prec$, one has $\configs^*\cap\configs(\prefixn_0^\prec)\ne\emptyset$.\eepr

\section{Discussion}
The results presented here give a toolkit for the analysis of tipping situations in a safe Petri net, i.e. when and how a basin boundary is crossed; an algorithmic method for finding minimally doomed configuration has been developed, implemented and tested. 

Moreover, we have introduced a measure of \emph{protectedness} that indicates the number of \emph{decisions} that separate a free state from doom. It uses an intrinsic notion of decisional height that allows to warn about impending dangerous scenarios; at the same time, this height is also 'natural' for unfoldings, in the sense that it induces an adequate linear order that allows to compute complete prefixes of bounded size.

On a more general level, the results here are part of a broader effort to provide a discrete, Petri-net based framework for dynamical systems analysis in the life sciences. The applications that we target lie in systems biology and ecology. 

Future work will investigate 
possibilities for  Doom Avoidance Control, i.e. devising strategies that allow to steer away from doom; we expect to complement the existing approaches via structural methods of e.g. Antsaklis et al \cite{IorAnt2006}, and also the unfolding construction of Giua and Xie~\cite{GiuaXie2005}.  A crucial question is the knowledge that any control player can be assumed to have, as a basis for chosing control actions. We believe the protectedness measure is a valid candidate for coding this information, so that a controller may take action when the system is too close to doom (wrt some thresholds to be calibrated) but there still remain decisns that can be taken to avoid it. 
Evaluating this option, along with other approaches, must, however, be left to future work.


\bibliographystyle{eptcs}
\bibliography{ecocontrol}  

\appendix
                            
\end{document}